\documentclass{ifacconf}

\usepackage{graphicx}      
\usepackage{natbib}        
\usepackage{bm}
\usepackage{mathtools,amsmath,amsfonts,amssymb}
\usepackage{color}
\usepackage{subfig}
\usepackage{xfrac}
\usepackage{algorithm}
\usepackage[noend]{algpseudocode}
\algnewcommand{\Initialize}[1]{%
	\State \textbf{Initialize:~$\forall j\in \mathcal{N}_i^{-}$}
	\Statex \hspace*{\algorithmicindent}\parbox[t]{.8\linewidth}{\raggedright #1}
}
\algnewcommand{\Ensured}[1]{%
	\State \textbf{Ensure:}
	\Statex \hspace*{\algorithmicindent}\parbox[t]{.8\linewidth}{\raggedright #1}
}

\newcommand{\sumij}{\sum_{\ell \in \mathcal{N}_i^{-}(t) \setminus j }\!\!\!\!\!\!}

\usepackage{tabularx}
\makeatletter
\newcommand{\multiline}[1]{%
	\begin{tabularx}{\dimexpr\linewidth-\ALG@thistlm}[t]{@{}X@{}}
		#1
	\end{tabularx}
}
\makeatother

\DeclareMathOperator*{\argmin}{arg\,min}

\begin{document}
\begin{frontmatter}

\title{Detection and Isolation of Adversaries in Decentralized Optimization for Non-Strongly Convex Objectives\thanksref{footnoteinfo}} 

\thanks[footnoteinfo]{This research was supported in part by Cybersecurity for Energy Delivery Systems program, of the U.S. Department of Energy, under contract DOE0000780 and the National Science Foundation CCF-BSF $1714672$ grant. Any opinions, and findings expressed in this material are those of the authors and do not necessarily reflect those of the sponsors.}

\author[First]{Nikhil Ravi} 
\author[First]{Anna Scaglione} 

\address[First]{School of Electrical, Computer and Energy Engineering,\\ Arizona State University, 
   Tempe, AZ 85281 USA \\(e-mail: \{Nikhil.Ravi,Anna.Scaglione\}@asu.edu).}

\begin{abstract}                
Decentralized optimization has found a significant utility in recent years, as a promising  technique to overcome the curse of dimensionality when dealing with large-scale inference and decision problems in big data. While these algorithms are resilient to node and link failures, they however, are not inherently Byzantine fault-tolerant towards insider data injection attacks. This paper proposes a decentralized robust subgradient push (RSGP) algorithm for detection and isolation of malicious nodes in the network for optimization non-strongly convex objectives. In the attack considered in this work, the malicious nodes follow the algorithmic protocols, but can alter their local functions arbitrarily. However, we show that in sufficiently structured problems, the method proposed is effective in revealing their presence. The algorithm isolates detected nodes from the regular nodes, thereby mitigating the ill-effects of malicious nodes. We also provide performance measures for the proposed method.
\end{abstract}


\end{frontmatter}

\section{Introduction}
A number of machine learning and big data problems can be generalized to take the following form:
\begin{equation}
\min_{\bm{x} \in \mathbb{R}^d} F(\bm{x}) = \min_{\bm{x} \in \mathbb{R}^d}\frac{1}{N}\sum_{i=1}^{N} f_i(\bm{x}),\label{eq:opt}
\end{equation}
where $F(\cdot): \mathbb{R}^d \rightarrow \mathbb{R}$ is the global objective which the nodes are trying to collectively minimize. However, the nodes only have access to their own private cost functions, $f_i(\cdot): \mathbb{R}^d \rightarrow \mathbb{R}$. There is a large class of decentralized peer-to-peer algorithms  for solving such \textit{decomposable} optimization problems via peer-to-peer consensus protocols, see~\cite{tsitsiklis1986distributed,nedic2009distributed} for early contributions and see~\cite{boyd2011distributed,nedic2018network} for extensive literature surveys.

While these algorithms are resilient to node or edge failures, they are not however inherently resilient to insider-based data injection attacks~(see e.g.~\cite{gentz2015detection,blanchard2017machine,gentz2016data,wu2018data,sundaram2018distributed}).
This has spurred significant interest on robust decentralized optimization algorithms. In literature, broadly speaking, there are two schools of solutions. In the first set of works, authors propose methods to regularize the objective by relaxing the hard consensus constraints, typically by using TV norm regularization scheme, see~\cite{ben-ameur2016robust,koppel2017proximity,xu2018robust}. 
Even if the adversarial environment can be studied similarly to the non-adversarial case, such approaches do not admit the same solutions as the original problem even when the attackers are not present at all. Also, their performance is highly dependent on the attack scenario, the choice of the regularizing function, and the regularizing parameter. 

These issues are not shared by the second set of works, where each node builds a score about their neighbors' updates and uses this score to detect their potential malicious intent, see~\cite{vukovic2014security,su2015byzantine1,su2015fault4,su2018finite,sundaram2018distributed,ravi2019adversarial}. These scores can then be used by regular nodes to dynamically sever ties with suspicious peers and continue to run the algorithm, see~\cite{ravi2019adversarial}. As long as the network remains strongly connected and set of the isolated nodes includes all the malicious nodes, the network converges to optimum solution of the objective that only includes the nodes that are not isolated.
This is possible for appropriate peer-to-peer optimization algorithms because they can restore normal operations by isolating the attackers while the algorithm is running.

The literature considers a number of attack models that differ based on the structure of the data injected by the malicious nodes. This paper focuses on attacks that involve coordinated malicious agents that follow algorithmic protocols while manipulating their private cost functions arbitrarily. 
We consider a network of nodes that are collectively solving problem~\eqref{eq:opt} by implementing the Push-Sum algorithm proposed in~\cite{kempe2003gossip}. 
Concurrently, the regular nodes maintain a \textit{maliciousness} score about their neighbors as a function of the local neighborhood data.
The idea is to sever ties with those deemed to be malicious and then continuing the updates according to Push-Sum.
This approach leverages the self-healing property of the gradient-based Push-Sum algorithm where nodes dynamically sever ties with those neighbors that are more likely to be malicious. The paper is organized as follows. In Section~\ref{sec:system_model}, the system model and the attack characteristics are described. In Section~\ref{sec:analysis}, we introduce the Robust Subgradient Push algorithm and its consequences. In Section~\ref{sec:detect_strats}, we present some simulation results that demonstrate the effectiveness of the proposed approach. We conclude in Section~\ref{sec:conclusion}.

\textbf{Notations}: Boldfaced lower-case (resp.\ upper-case) letters denote vectors (resp.\ matrices) and  
$x_i$ ($X_{ij}$) denotes the $i$th element of a vector $\bm{x}$ (the $ij$th entry of a matrix $\bm{X}$).
Calligraphic letters are sets and $|\mathcal{A}|$ denotes the cardinality of a set $\mathcal{A}$; difference between two sets $\mathcal{A}$ and $\mathcal{B}$ is denoted by $\mathcal{A}\setminus \mathcal{B}$.

\section{System Model}\label{sec:system_model}
The set of nodes in the system are modeled by a directed graph $\mathcal{G} (\mathcal{V},\mathcal{E})$, where $\mathcal{V}$ is the set of nodes and $\mathcal{E}$ is the set of edges that represent the communication links between the nodes. To describe the adversarial environment, the set $ \mathcal{V} $ is partitioned into two subsets, namely: the set
$\mathcal{V}_r $ of regular nodes (RNs) with $ N_r :=|\mathcal{V}_r|$, and the set $\mathcal{V}_m $ of malicious nodes (MNs) 
with $N_m :=|\mathcal{V}_m|$, so that $ \mathcal{V} = \mathcal{V}_r \cup \mathcal{V}_m$. 
Similarly, the edge set $\mathcal{E}$ is also partitioned $\mathcal{E} = \mathcal{E}_r \cup \mathcal{E}_m$, 
where $ \mathcal{E}_r $ is the set of links emanating from RNs, 
while $\mathcal{E}_m$ is the set of links emanating from the MNs. 
The set $\mathcal{E}_r$ is further partitioned $\mathcal{E}_m = \mathcal{E}_{mr} \cup \mathcal{E}_{mm}$, 
while $\mathcal{E}_{mm}$ is the set of links emanating and ending at MNs. The goal of the RNs is to identify and sever all the edges belonging to $\mathcal{E}_{mr}$. Let $\mathcal{N}_i^{-}$ be the set of all nodes that can send information to node $i$ and $\mathcal{N}_i^{+}$ be the set of all nodes that can receive information from node $i$.

\textbf{Problem Statement}: In the presence of adversaries in a network, the benchmark for th performance is the analogous problem as in~\eqref{eq:opt}, where the graph is replaced with the graph formed by the regular nodes only,
i.e., 
\begin{equation}\label{eq:adver-opt}
\begin{aligned}
& \underset{\{\bm{x_i, i\in \mathcal{V}_r}\}}{\text{minimize}}~\frac{1}{|\mathcal{V}_r|} \sum_{i \in \mathcal{V}_r} f_i(\bm{x}_i)
~~\text{s.t.}~~
\bm{x}_{i}=\bm{x}_{j}\;\forall~ij \in \mathcal{E}_{rr},
\end{aligned}
\end{equation}
where $f_i: \mathbb{R}^d \rightarrow \mathbb{R}$ is the private cost/loss function only available at node $i$ and the constraint enforces consensus in the network's neighborhoods. Decentralized  optimization algorithms designed for arbitrary network topologies require doubly-stochastic weight matrices. Techniques such as Subgradient-Push (SGP) relax this requirement \cite{nedic2018network}. Further improvements in the convergence rates algorithms for directed graphs are presented in DEXTRA (\cite{xi2017dextra}) and  ADD-OPT/Push-DIGing (\cite{xi2018addopt}).  
However,  these  algorithms  require each node to know its out-degrees (the number of its out-neighbors) which  is  not  a  realistic  assumption in general, and is even more problematic in an adversarial environment. 

Before moving on to describe the characterization of malicious nodes, we introduce a few assumptions.
 \begin{assum}{(Strong Connectivity)}\label{as:connectivity} 
 	The sequence of graphs, $\{\mathcal{G}(t)\}$, induced by the dynamic in- and out-neighborhoods over time is strongly connected.
 \end{assum}
 \vspace{-0.5em}
\begin{assum}{(Convexity)}\label{as:convexity} 
	Each of the functions $f_i, \forall i \in \mathcal{V}$, is convex over the $\mathcal{R}^d$.
\end{assum}
 \vspace{-0.5em}
\begin{assum}{(Bounded Subgradients)}\label{as:bounded_subgradients} 
	All the subgradients of each of the functions $f_i,\forall i \in \mathcal{V}$, are bounded. That is, $\exists~ l_i < \infty$ such that $\|\bm{g}_i\|\leq l_i,\forall i \in \mathcal{V}$, for all subgradients $\bm{g}_i$ of $f_i(\bm{x})$ and for all $\bm{x}\in\mathbb{R}^d$.
\end{assum}
 \vspace{-0.5em}
\begin{assum}{(Malicious nodes)}\label{as:malicious_nodes} 
	Let us suppose that the sequence of graphs $\{\mathcal{G}(t)\}$ is $(2\kappa+1)$-strongly connected. Then there exists at most $\kappa$ adversaries in the whole network, i.e., $|\mathcal{V}_m| \leq \kappa$. This also implies that there exists at most $\kappa$ adversaries in a node's in-neighborhood, i.e., $|\mathcal{N}_i^{-}\cup \mathcal{V}_m| \leq \kappa,\forall i \in \mathcal{V}$.
\end{assum}

If assumptions~\ref{as:connectivity}-\ref{as:bounded_subgradients} hold and under no interference from malicious nodes, the convergence of the original Gradient-Push was proven in~\cite{nedic2015distributed}, i.e.,
\[
{\bm x}_i^{\infty}\triangleq\lim_{t\rightarrow\infty} \bm{x}_i(t) = \bm{x}^* = \argmin_{\bm{x} \in \mathbb{R}^d} F(\bm{x}), ~~\forall~~i\in {\cal V} 
\]
\subsection{Attack characterization} 
Let $F^r(\bm{x})$ be the cost and $\bm{x}^*$ the solution of  \eqref{eq:adver-opt}, i.e.:
\begin{align}
F^r(\bm{x})\triangleq \frac{1}{|\mathcal{V}_r|} \sum_{i \in \mathcal{V}_r} f_i(\bm{x})~,~~   
\bm{x}^* \triangleq \argmin_{\bm{x}} \sum_{i\in {\mathcal V}_r}f_i(\bm{x})\label{eq:sol_with_adv}.
\end{align}
The performance of any algorithm that is resilient towards an attack can be measured using several  metrics that capture the deviation of the limiting state
from the ideal asymptotic condition in \eqref{eq:sol_with_adv}, for example:
\begin{itemize}
    \item {\bf Average solution difference}:
    \begin{equation}\label{eq:ep_p}
        \epsilon_{p}\triangleq
        \frac{1}{|\mathcal{V}_r|} \sum_{i \in \mathcal{V}_r} \|{\bm x}_i^{\infty}-{\bm x}^*\|_{p}
    \end{equation}
    \item {\bf Average cost increase}:
    \begin{equation}\label{eq:func_err}
        \varrho\triangleq
        \frac{1}{|\mathcal{V}_r|} \sum_{i \in \mathcal{V}_r} f_i({\bm x}_i^{\infty})-F^r({\bm x}^*)
    \end{equation}
    \item {\bf Average deviation from consensus}:
      \begin{equation}
        \gamma\triangleq\frac{1}{|\mathcal{V}_r|}\sum_{i \in \mathcal{V}_r}\left\|{\bm x}_i^{\infty}-\overline{\bm x}^{\infty}  
        \right\|,~~
\overline{\bm x}^{\infty}   \triangleq     \frac{1}{|\mathcal{V}_r|} \sum_{\ell \in \mathcal{V}_r} {\bm x}_{\ell}^{\infty}
    \end{equation}
\end{itemize}
\begin{figure}[htbp!]
	\centering
	\includegraphics[scale=0.5]{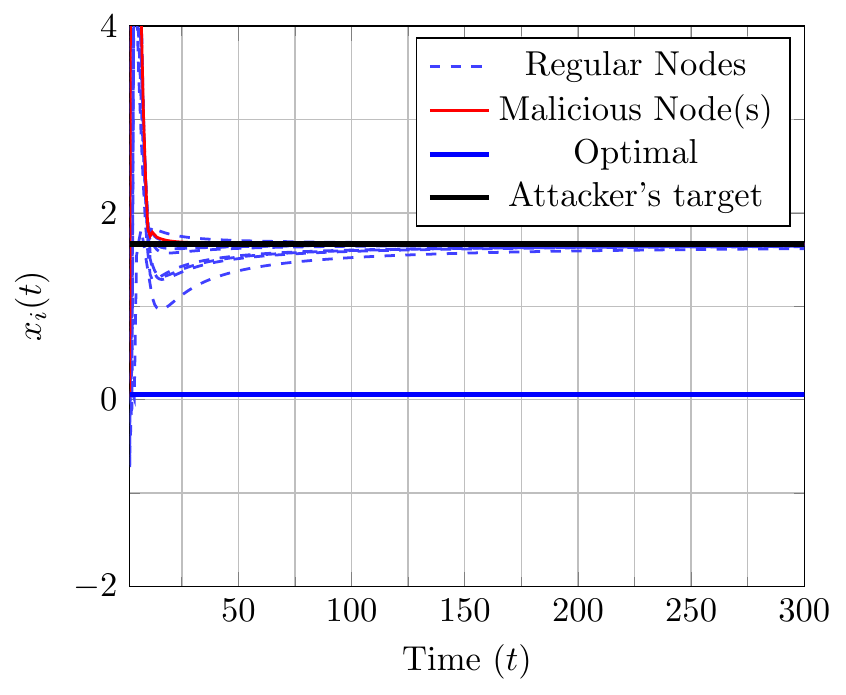}
	\caption{Consensus without attacker detection with adversaries following algorithmic protocols but modifying their local cost functions dynamically to reach their intended target (in black). Here, the regular nodes in blue converge to the attacker's (in red) target.
	In the example a set of nodes ({$1-10$}) are trying to solve problem~\ref{eq:opt} using the standard Sub-gradient push (SGP) in a static network. However, a malicious node drives the network towards its target, $x^a$. 
}
	\label{fig:astray}
\end{figure}

It is a well established that the presence of even a single malicious node will lead the convergence of the algorithm astray (see e.g. Figure~\ref{fig:astray}). 
The type of attack that is considered in our paper, represents the case where the regular nodes' functions $f_i({\bm x}),~i~\in~{\cal V}_r$ have common statistical characteristics. 
The 
attacker follows the algorithm as well, however, its intent is to alter the solution, which requires altering the statistics of the function and its gradient. For example, this situation arises when the distributed optimization objective is to solve a regression problem where, in normal conditions, the sensors cooperatively are trying to estimate a vector ${\bm x}_o$ based on a noisy observation model: 
\begin{equation}
    s_i=h_i({\bm x}_o)+w_i~~\in~{\mathbb{R}}
\end{equation}
for all $i\in {\cal V}$ where $w_i$ are independent identically distributed noise samples. 
The regression can be formulated as the minimization of the sum of $f_i({\bm x})=\|s_i-h_i({\bm x})\|^2$, which amounts to seeking the maximum ${\bm x}$. To be concrete, we will focus on Gaussian independent identically distributed (i.i.d.) noise case $w_i\sim {\cal N}(0,\sigma^2)$ in which case:
\begin{equation}
    f_i({\bm x})=\|s_i-h_i({\bm x})\|^2
\end{equation}
and the gradient\footnote{Note that in this case, given that the noise terms are non bounded, the condition stated as Assumption 3 of bounded gradients can only be guaranteed with high probability. To avoid complications in the discussion we will discuss convergence as if the condition is met. } 
is:
\begin{equation}
    {\bm g}_i({\bm x})=2\nabla h_i({\bm x})(s_i-h_i({\bm x}))
\end{equation}
The attack can be caused by a spoofing attack to a set of sensors:
\begin{equation}
    s^a_m=s_m+\delta s_m~~~m\in~{\cal V}_m
\end{equation}
Under this attack, the gradient of the functions 
$f_m^a({\bm x})=\|s_m^a-h_m({\bm x})\|^2$ is, therefore perturbed:
\begin{align}
\!\!{\bm g}_m({\bm x})=&2\nabla h_m({\bm x})(s^a_m-h_m({\bm x}))=  {\bm g}_m({\bm x})+\delta{\bm g}_m({\bm x})
\end{align}
Under such an attack, a strongly connected  network will converge to consensus on the following solution:
\begin{equation}\label{eq:with_adv}
{\bm x}^a=\argmin_{{\bm x}}\left(\frac{|{\cal V}_r|}{|{\cal V}|}F^r({\bm x})+
\frac{1}{|{\cal V}|}\sum_{i\in {\cal V}_m}
f_i^a({\bm x})\right)
\end{equation}
with average deviation from consensus $\gamma=0$ and average solution difference and cost deviation, respectively:
\begin{equation}
\epsilon_p=\|{\bm x}^a-{\bm x}^*\|_p,~~\varrho=F^r({\bm x}^a)-F^r({\bm x})    
\end{equation}

In the context of this type of attack, the percentage increase the additional average error relative to the latent parameter ${\bm x}_o$  measures the damage caused by the malicious agents as well, that is for instance:  
\begin{equation}\label{eq:degradation}
\xi_p\triangleq\frac{
\frac{1}{|{\cal V}|} \sum_{i\in {\cal V}} 
\|\bm{x}^{\infty}_i -\bm{x}_o\|_p
}{\|\bm{x}^* -\bm{x}_o\|_p}
\end{equation}
where a significant damage comes from a large value of $\xi_p$ in the presence of an attack; since when the attack is fully successful there is consensus on $\bm{x}^a$ the value of $\xi_p$ is:
\begin{equation}\label{eq:degradation2}
\xi_p=\sfrac{\|\bm{x}^a -\bm{x}_o\|_p}{\|\bm{x}^* -\bm{x}_o\|_p}
\end{equation}

Of course, the quantity that we consider a benchmark $\|\bm{x}^* -\bm{x}_o\|_p$ itself represents a loss relative to the case where no sensor is attacked, as more observations improve the estimation error performance. 

In our previous work \cite{ravi2019adversarial} we noted that staging a sufficiently strong attack in terms of $\epsilon_p$, requires either sufficiently large gradients for the malicious nodes, or a special choice for the target point ${\bm x}^a$. In fact, since $\bm{x}^a$ is the stationary point for the problem in \eqref{eq:with_adv}, we have
\[
\textstyle{\sum}_{i\in {\mathcal V}_r} \bm{g}_i (\bm{x}^a)+ \textstyle{\sum}_{i\in {\mathcal V}_m} \bm{g}_i(\bm{x}^a)= \bm{0}.\]
Let ${\bm H}_i({\bm x})$ be the Hessian of the function $f_i({\bm x})$ and ${\bm H}^r({\bm x})$ the Hessian of the function $F^r({\bm x})$; then:
\begin{align}
\textstyle{\sum}_{i\in \mathcal{V}_m} \bm{g}_i (\bm{x}^a) &\approx - \textstyle{\sum}_{i\in \mathcal{V}_r} \bm{g}_i (\bm{x}^*)+{\bm H}_i(\bm{x}^*)(\bm{x}^a-\bm{x}^*)\\
&=-
{\bm H}^r(\bm{x}^*)(\bm{x}^a-\bm{x}^*).
\end{align}
The last equation implies that the most effective attacks can be staged when the malicious nodes can be selected so that the Hessian ${\bm H}^r(\bm{x}^*)$ is singular, compromising the convergence of the algorithm even if the malicious nodes were isolated and the network of regular nodes remains strongly connected. 

This requires us to state the following assumption:
\begin{assum}\label{assu-5}
The Hessian ${\bm H}^r(\bm{x}^*)$ of $F^r({\bm x})$ is positive definite. 
\end{assum}
\begin{prop}[\cite{ravi2019adversarial}]
Under Assumption \ref{assu-5}, the nodes achieve consensus (i.e. $\gamma=0$), and the average solution difference is such that:
\begin{equation}\label{eq-bound}
\left\|\textstyle{\sum}_{i\in \mathcal{V}_m} \bm{g}_i (\bm{x}^a)\right\|^2_2\gtrsim\epsilon_2
\lambda_{\min}({\bm H}^r(\bm{x}^*)).   
\end{equation}
\end{prop}

\section{Robust Subgradient Push}\label{sec:analysis}
The basic idea of this paper is to leverage the structure of the problem, as well as the  need of the malicious agents to use large gradients to influence the result, as it is made apparent by equation \eqref{eq-bound}.
We wish to use this fact to design a score $S_{ij}(t)$ that each node updates $\forall j\in{\cal N}_i$ and uses to detect and then reject updates from a specific node. 

In this paper we propose a modified version of the Subgradient-Push (\cite{nedic2015distributed}) that is based on the original Push sum algorithm (\cite{kempe2003gossip}), see algorithm~\ref{alg:sub_push}. 
This algorithm differs from the original Subgradient-Push in the fact that at all times $t>0$, the out-degree of all the nodes is $2$, i.e., instead of a broadcast communication model, each node chooses one of its out-neighbors uniformly at random and sends its information to that node and to itself. This eliminates the necessity of each node knowing its out-degree, the trade-off being a slower convergence rate. The algorithm converges even when the functions are non strongly convex, which is the typical scenario of interest in a network of sensors that collect a scalar measurement while the goal is to estimate a $d$-dimensional vector parameter ${\bm x}_o$.

We name it Robust Subgradient-Push (RSGP). In RSGP
each node $i$ maintains a set of variables, $\bm{z}_i(t), \bm{v}_i(t) \in \mathbb{R}^d$ and $y_i(t) \in \mathbb{R}$, at all times $t\geq 0$. The algorithm starts with an arbitrarily initiated $x_i(0) \in \mathbb{R}^d$, and with $y_i(0)=1$ at each node $i$, and is descried in Algorithm~\ref{alg:sub_push}.
\begin{algorithm}[htbp!]
	\caption{Robust Subgradient-Push (RSGP)}\label{alg:sub_push}
	\begin{algorithmic}[1]
		\Initialize{$\bm{z}_i(0) \in \mathbb{R}^d$ arbitrarily, $y_i(0) = 1$, $\tilde{{S}}_{ij}(0) = \bm{0}$}
		\Ensured{$\{\eta\}_t$ satisfies condition~\eqref{eq:step_size} and
			$\alpha < 1$}
		\For {$t\geq 0$}
		\For {$i \in \mathcal{V}$}
		\State \multiline{Let $\{{\bm{z}}_j(t),{y}_j(t)\}, \forall j \!\in\! \mathcal{N}_i^{-}(t)\! \subseteq \! \mathcal{N}_i^{-}$, be the pairs of data sent to node $i$ at time $t$}
		\State \multiline{Send $\{\bm{z}_i(t),y_i(t)\}$ to $i$ itself and a node $k \in \mathcal{N}_i^{+}$ chosen uniformly at random; $\mathcal{N}_i^{+}(t) = \{i,k\}$}
		\State Update $\bm{v}_i,y_i,\bm{x}_i$ and $\bm{z}_i$ as follows:
		\begin{align}
	        \bm{v}_i(t+1) &= \frac{1}{2}\sum_{j\in \mathcal{N}_i^{-}(t)} {\bm{z}}_j(t)\\
		y_i(t+1) &= \frac{1}{2}\sum_{j\in \mathcal{N}_i^{-}(t)} {y}_j(t)\\
		\bm{x}_i(t+1) &= \bm{v}_i(t+1)/y_i(t+1)\\
		\bm{z}_i(t+1) &= \bm{v}_i(t+1) - \eta_{t+1}\bm{g}_i(t+1)
		\end{align}
		\If {$i \in \mathcal{V}_r$}
		\ForAll {$j \in \mathcal{N}_i^{-}$}
		\State \multiline{Calculate the score $\tilde{S}_{ij}(t)$ from  equation~
		\eqref{eq:avg_score}.}
		\EndFor
		\State Calculate $\chi_i(t)$ from equation~\eqref{eq:thresh}.
		\ForAll {$j \in \mathcal{N}_i^{-}$}
		\If {$\tilde{S}_{ij}(t) > \chi(t)$}
		\State $\mathcal{N}_i^{-} \gets \mathcal{N}_i^{-}\setminus j$ and  $\mathcal{N}_i^{+} \gets \mathcal{N}_i^{+}\setminus j$
		\EndIf
		\EndFor
		\EndIf
		\EndFor
		\EndFor
	\end{algorithmic}
\end{algorithm}
Here, $\bm{g}_i(t) = \nabla f_i(\bm{x}_i(t))$ is a subgradient of $f_i(\cdot)$ calculated at $\bm{x}_i(t)$, and $\eta_t>0$ is the step-size that decays with time such that 
\begin{equation}\label{eq:step_size}
\textstyle{\sum}_t \eta_t = \infty, ~~~ \textstyle{\sum}_t \eta^2_t < \infty.
\end{equation}

\subsection{Detection and Isolation via a Neighborhood Score}

Crucial to RSGP is the definition of the score that shall reveal the malicious nodes. 
We note that each node receives its neighbors' $\bm{z}_i(t)$ and $y_i(t)$ state variables when they communicate. The variable $y_i(t)$ of node $i$ is a function of the number of instantaneous in-neighbors $i$ receives at any particular time instant. Importantly, it does not depend on node $i$'s gradient, which is what the sensor spoofing attack manipulates. The iterative update of variable $\bm{z}_i(t)$, however, includes a gradient step. Thus, we define a \textit{maliciousness} score $S_{ij}(t)$ that can be maintained by each regular node $i$ about their neighbors $\mathcal{N}_i^{-}$. 
As the nodes approach consensus, their direction of descent is typically dominated by the malicious nodes' gradients. The intuition is that node $i$ can use:
\begin{equation}
\hat{\bm{x}}_j(t) :=  {\bm{z}}_j(t)/{y}_j(t)
\end{equation}
to track neighbor $j$'s instantaneous gradient.
 From the bound in Equation~\eqref{eq-bound}, we hypothesize that the malicious nodes will appear as outliers relative to the rest of the nodes in the neighborhood, as long as they are sufficiently outnumbered. 
The instantaneous score we propose is: 
\begin{equation}\label{eq:inst_score}
S_{ij}(t) \triangleq \frac{1}{\eta^2_t}\left\Vert \sumij \left(\hat{\bm{x}}_j(t)  - \hat{\bm{x}}_\ell(t) \right) \right\Vert^2_2.
\end{equation}
To gain some intuition, suppose the nodes are approaching consensus. At this point the values of $y_i(t)$ are likely to have converged as well, and:
$$
{\bm x}_j(t)={\bm v}_j(t)/y_j(t)\approx {\bm x}^a
$$
In this case:
\begin{align}
\hat{\bm x}_j(t+1)-\hat{\bm x}_{\ell}(t+1)&= \frac{{\bm v}_j(t)}{y_j(t)}-
\frac{{\bm v}_{\ell}(t)}{y_{\ell}(t)}-
\eta_t\left(
{\bm g}_j(t)-{\bm g}_{\ell}(t)
\right)\nonumber\\
&\approx \eta_t\left({\bm g}_{\ell}(t)-
{\bm g}_j(t)\right), \label{eq.approxS}
\end{align}
which indicates that the score is essentially comparing the disparity in the gradients and:
\begin{align}
S_{ij}(t)&\approx
\left\Vert \sumij \left({\bm{g}}_{\ell}(t)  - {\bm{g}}_j(t) \right) \right\Vert^2_2\\
&=
\left\Vert \sumij  {\bm{g}}_{\ell}(t)   - (d_i(t)-1){\bm{g}}_j(t) \right\Vert^2_2.    \label{eq:g}
\end{align}
Let us consider for simplicity the presence of only one malicious agent in the $i$th node neighborhood. 
 In the following, we denote node $i$ degree by $d_i(t)=|{\cal N}_i(t)|$.
The last equation, together with the observation in Proposition 1, implies that the score for a malicious neighbor $m$, for $t$ sufficiently large is approximately: 
\begin{align}
S_{im}(t)&\approx (d_i(t)-1)^2\Vert{\bm{g}}_m(t) \Vert^2_2.    
\end{align}
On the other hand, it is not difficult to show expanding the norm squared in \eqref{eq:g} that:
\begin{align}
    \frac{1}{d_i}\!\sum_{i\in {\cal N}_i(t)}\!\!\!\!
    S_{ij}(t)\!&= d_i(t)\!\!
    \sum_{i\in {\cal N}_i(t)}\!\!\!\|{\bm g}_i(t)\|^2+\\
    &-\left(2-\frac{1}{d_i(t)}\right)\left\Vert\sum_{i\in {\cal N}_i(t)}\!\!\!
    {\bm g}_i(t)\right\Vert^2\!
    \lessapprox d_i(t)\Vert{\bm{g}}_m(t) \Vert^2_2\nonumber
\end{align}
which suggests that, as long as $(d_i(t)-1)^2>d_i(t)$ (which happens as long as $d_i(t)\geq 4$) the malicious agent score will tend to dominate the average of the scores of the neighborhood.
Considering the randomness that exist in the protocol iterations, the score used to detect the agents is averaged over time. In fact, RSGP uses a  cumulative score is given by:
\begin{equation}\label{eq:avg_score}
\tilde{S}_{ij}(t) = \sum_{\tau = 1}^{t} \alpha^{\tau}S_{ij}(\tau)=\tilde{S}_{ij}(t-1)+ \alpha^{t}S_{ij}(t),
\end{equation}
where $0<\alpha<1$. The weighted average is designed in such a manner that older instantaneous scores are given lower weights than newer ones, for which \eqref{eq.approxS} is more accurate.

Regular Nodes can then dynamically isolate those neighbors whose score cross a certain threshold, $\chi_i(t)$.
Let $\tilde{\bm S}_{i}(t)$ be the vector of scores at node $i$. In our implementation we choose $\chi_i(t)$ as follows:
\begin{equation}\label{eq:thresh}
\chi_i(t) = 
\mathrm{avg}\left(\tilde{\bm S}_{i}(t)\right)+\beta\times
\mathrm{std}\left(\tilde{\bm S}_{i}(t)\right)
\end{equation}
where $\mathrm{avg}(\bm{a})$ and $\mathrm{std}(\bm{a})$ are sample average and sample standard deviation of vector $\bm{a}$ entries. The parameter $\beta$ controls the aggressiveness of the edge severing strategy.



It is important to note that as a result of edge severing, depending on the aggressiveness of the isolation strategy, some regular nodes might also get isolated from the rest of the regular nodes. This may result as a consequence of the regular node being slow in isolating neighboring malicious node(s). In another scenario, the algorithm might give rise to a splintering in the network to multiple connected subgraphs $\overline{\mathcal{G}}_\ell$, thereby leading to polarities in the final convergence points of such subgraphs to $\bm{x}^\infty_\ell$. Ideally, the RSGP at the optimum value of $\beta$ separates $\mathcal{G}$ into $\mathcal{G}_r$ made up solely of the regular nodes, and if the malicious nodes are coordinating (which they are in this paper), into $\mathcal{G}_m$ made up solely of the malicious nodes. If the malicious nodes are not cooperating, then we may see further splintering in $\mathcal{G}_m$.
The various scenarios and their error plots are discussed in section~\ref{sec:detect_strats}.

\section{Numerical Results with a case study}\label{sec:detect_strats}

To illustrate our approach, let us consider a parameter estimation problem where the observation model is given by $s_i = \bm{h}_i^\text{T}\bm{x}_o + {w}_i$, where $\bm{h}_i\in\mathbb{R}^d$, ${w}_i\in\mathbb{R}$ represents i.i.d noise samples drawn from a normal distribution with mean zero and variance $\sigma^2_i$, for all $i \in \mathcal{V}$. Then the local linear least square loss function may be written as:
\[
f_i(\bm{x}) = (\bm{h}_i^\text{T}\bm{x}_i - s_i)^2.
\]
The global cost function is $
F(\bm{x}) = (\sfrac{1}{|{\cal V}|})\sum_{i\in\mathcal{V}}f_i(\bm{x})$ 
when all the nodes are performing regularly. However, when certain nodes are attacked or are malicious, the regular nodes estimate $\bm{x}^*$ by solving the following problem
\[
\min_{\bm{x}} F^r(\bm{x}) = \min_{\bm{x}} \frac{1}{|\mathcal{V}_r|}\sum_{i\in\mathcal{V}_r}(\bm{h}_i^\text{T}\bm{x}_i - s_i)^2.
\]
Let us assume without loss of generality that ${\cal V}_r=\{1,\ldots,N_r\}$ and define the matrix $
{\bm A}_r^\text{T}=\begin{bmatrix}{\bm h}_1,
\ldots,{\bm h}_{N_r}\end{bmatrix}.  
$
Denoting by ${\bm A}_r^{\dagger}:=\left(\bm{A}_r^\text{T}\bm{A}_r\right)^{-1}\bm{A}_r^\text{T}$, which is the pseudoinverse of matrix ${\bm A}_r$, the minimizer for this problem is known in closed-form and given by 
\[
    {\bm{x}}^* = {\bm A}_r^{\dagger}\bm{s}_r = \bm{x}_o + \left(\bm{A}_r^\text{T}\bm{A}_r\right)^{-1}\bm{A}_r^\text{T}\bm{w}_r.
\] 
and $\bm{A}_r^\text{T}\bm{A}_r$ is the Hessian, which under Assumption \ref{assu-5} is full rank.  The cost is:
\begin{equation}
F^r({\bm{x}}^*)=\|
({\bm I}-{\bm A}_r^{\dagger}{\bm A}_r){\bm s}\|^2=
\|
({\bm I}-{\bm A}_r^{\dagger}{\bm A}_r){\bm w}\|^2
\end{equation}

Now, let $\bm{x}^\infty_i, \forall i \in \mathcal{V}_r$, be the points at which the regular nodes converge at the end of RSGP. The average solution error with respect to $\bm{x}^*$ is given by $\epsilon_p$ from equation~\eqref{eq:ep_p}.
The average cost increase of RSGP is given by $\varrho$ from equation~\eqref{eq:func_err}.
We can also define a regret in the errors as $\xi_p$ from equation~\eqref{eq:degradation}, which compares the error under attack and the error without an attack. We redefine the terms for ease of reading:
\begin{align*}
\epsilon_{p}&\triangleq \frac{1}{N_r} \sum_{i \in \mathcal{V}_r} \|{\bm x}_i^{\infty}-{\bm x}^*\|_{p},~~\varrho\triangleq \frac{1}{N_r} \sum_{i \in \mathcal{V}_r} \left(f_i({\bm x}_i^{\infty})-f_i({\bm x}^*)\right)%
\\
\gamma_p&\triangleq\frac{1}{N_r}\sum_{i \in \mathcal{V}_r}\left\|{\bm x}_i^{\infty}-\overline{\bm x}^{\infty}  \right\|_p,%
~~
\xi_p\triangleq\frac{\frac{1}{N_r} \sum_{i\in {\cal V}_r} \|\bm{x}^{\infty}_i -\bm{x}_o\|_p}{\|\bm{x}^* -\bm{x}_o\|_p},
\end{align*}
where $\overline{\bm x}^{\infty}   \triangleq     \frac{1}{|\mathcal{V}_r|} \sum_{\ell \in \mathcal{V}_r} {\bm x}_{\ell}^{\infty}$ and $p=2$.

\begin{figure}[htbp!]
	\centering
	\subfloat[]{%
	\label{fig:network}%
	\includegraphics[width=0.2\textwidth]{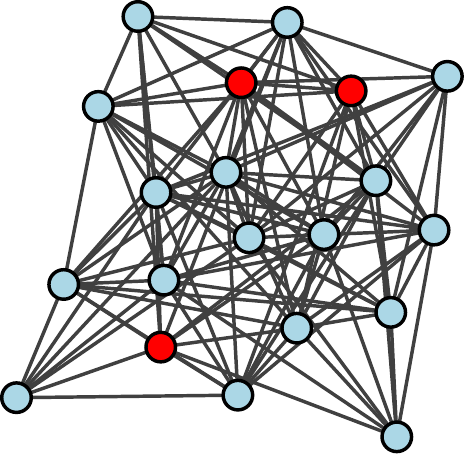}%
	}%
	\quad
	\subfloat[]{%
	\label{fig:convergence}%
	\includegraphics[width=0.22\textwidth]{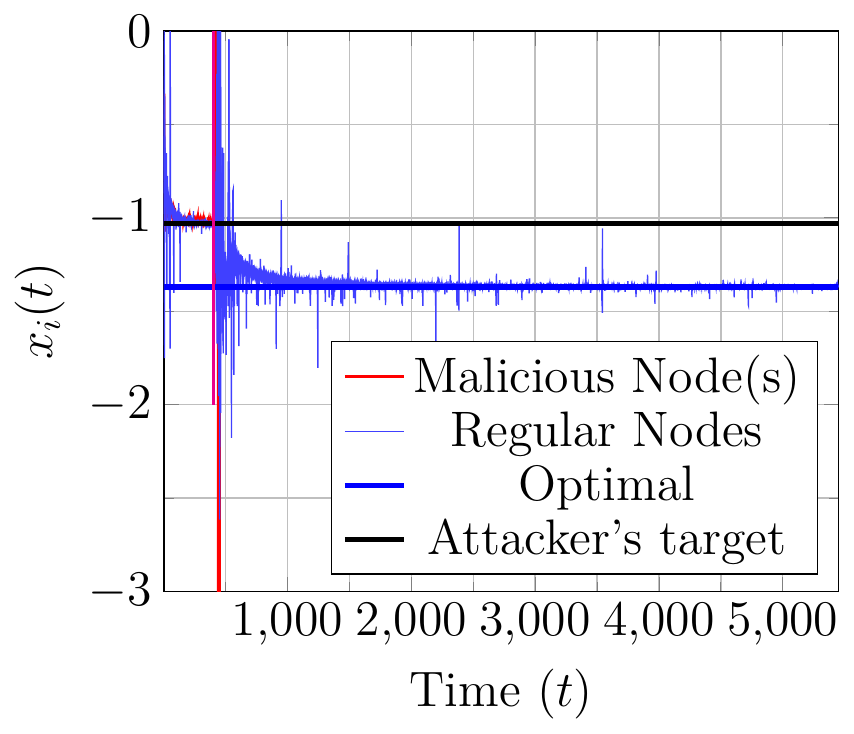}%
	}%
	\\
	\subfloat[]{%
	\label{fig:residual}%
	\includegraphics[width=0.22\textwidth]{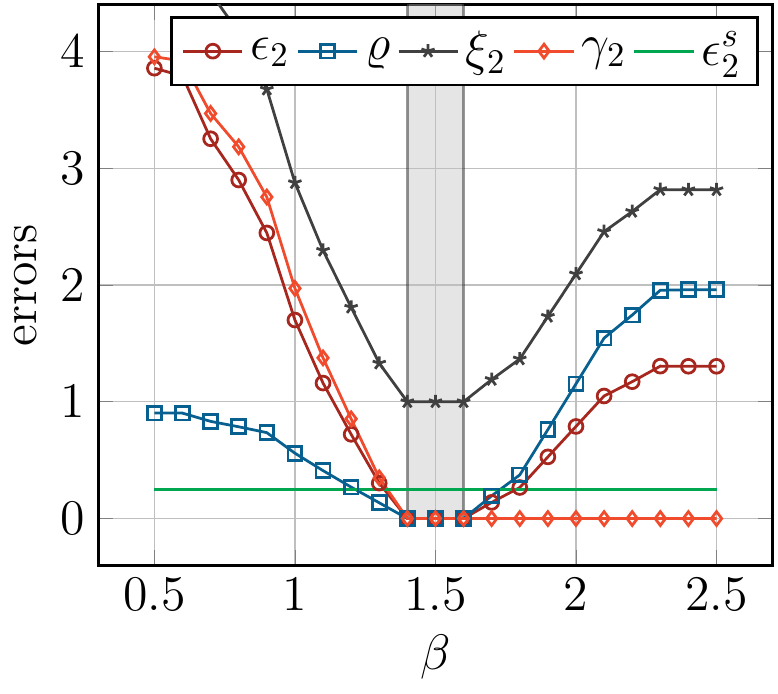}%
	}%
	\quad
	\subfloat[]{%
	\label{fig:total_var}%
	\includegraphics[width=0.22\textwidth]{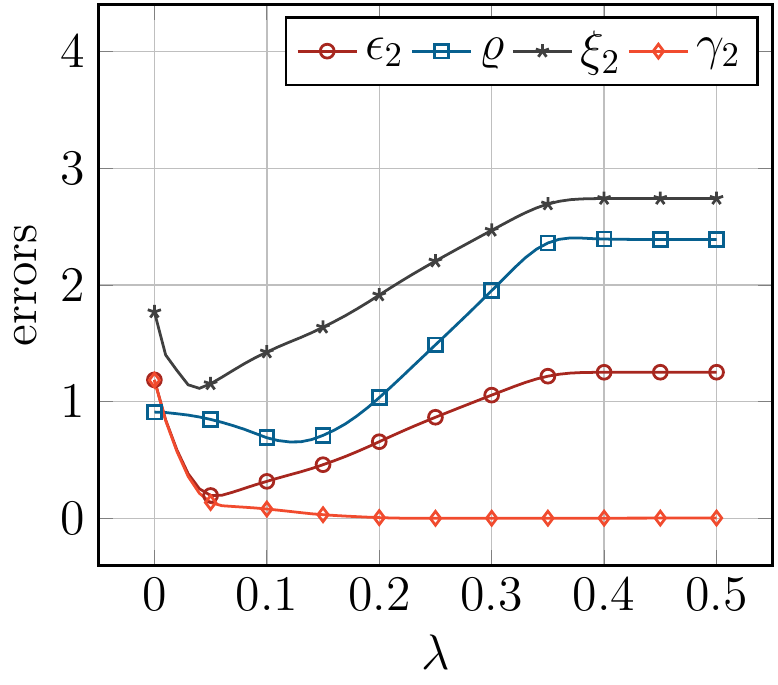}%
	}%
	\caption{\protect\subref{fig:network} Erd\H{o}s--R\'enyi network under consideration. \protect\subref{fig:convergence} Convergence to optimal consensus point $x^*$. \protect\subref{fig:residual} Average Residual as a function of $\beta$; $\epsilon^s_2$ is the solution difference for the algorithm in~\cite{sundaram2018distributed}. \protect\subref{fig:total_var} Average Residual as a function of $\lambda$ for the method in~\cite{ben-ameur2016robust}.}
	\label{fig:numericals}
\end{figure}

The figures in Figure~\ref{fig:numericals} show the performance of RSGP and compares it with the methods proposed in~\cite{ben-ameur2016robust} and~\cite{sundaram2018distributed}. The system setup is as follows.  We consider multiple realization of Erd\H{o}s--R\'enyi networks with $N=20$ nodes and $p=3\log(N)/N$. Without loss of generality, the node sets $\mathcal{V}_r=\{1,\ldots,17\}$ and $\mathcal{V}_m = \{18,19,20\}$ are kept the same for every montecarlo trial, but the edge set $\mathcal{E}$ is changed.
The vector $\bm{z}_i\sim \mathcal{N}(\bm{0},\bm{I})\in \mathbb{R}^d ~(d=2)$ is drawn randomly at the start of each montecarlo trial for all $i \in \mathcal{V}$. The following parameters/system variables remain the same for all the montecarlo trials; $\bm{x}_o \sim \mathcal{N}(\bm{0},\bm{I}) \in \mathbb{R}^d$, $\bm{h}_i\sim \mathcal{N}(\bm{0,\sigma^2\bm{I}}) \in \mathbb{R}^d$, $\bm{s}_i = \bm{h}_i^{\text{T}}\bm{x}_o + \bm{w}_i$, $\bm{w}_i\sim \mathcal{N}(0,1),~\forall i \in \mathcal{V}_r$. Note that $\bm{x}_o$ was initialized to $[0.0859,-1.4916]^\text{T}$, and the malicious nodes ($i \in \mathcal{V}_m$) alter their $\bm{h}_i$ as $\left((1/N_r)\bm{1}^\text{T}\bm{A}^r - 5\right)$ and ${s}_i$ as $\left((1/N_r)\bm{1}^\text{T}\bm{s}^r + 5\right)$.
One such realization is shown in figure~\ref{fig:network}.
In this network, nodes in color {\color{red}red} are attacking the network by altering the parameters of its loss function $f_i,~\forall i \in \mathcal{V}_m$. Let us suppose that the regular nodes keep a track of the metric in equation~\eqref{eq:avg_score} over time. 
Each node isolates those neighbors that exceed the threshold in equation~\eqref{eq:thresh} with $\beta= 1.5$. Figure~\ref{fig:convergence} shows the convergence of RSGP. At time $t=401$ (indicated in {\color{magenta}magenta}), all the nodes with the malicious node(s) in their neighborhood have successfully isolated the malicious agents. 
We then see the self-healing property of distributed consensus algorithm come into play and the regular nodes converge at the optimal consensus point in {\color{blue}\textbf{blue}}.

In Figure~\ref{fig:residual}, we plot $\epsilon_2, \varrho$, $\xi_2$, and $\gamma_p$ as functions of $\beta$, averaged over multiple montecarlo trials for realizations of $\mathcal{G}$. We see that at lower values of $\beta$, where nodes sever ties rather aggressively, the network is broken into multiple strongly connected subgraphs or into a completely disconnected network. This results in errors greater than zero. Whereas, at the intermediate values of $\beta$, where the regular nodes sever ties in a relatively passive manner, RSGP isolates almost all, if not all, the malicious nodes in the network and thus, we see a decrease in the errors. At the optimum choice for $\beta$ ($\beta^*\in(1.4,1.6)$ in this setup), the RSGP algorithm almost certainly disconnects the malicious nodes from the regular nodes and thus the errors reduce to zero. However, as $\beta$ increases past this region, the nodes start to become highly conservative in their cutting, leaving almost all the malicious nodes still connected to the network. Thus, we see an increase in the errors and after a certain value of $\beta$, when no edge cutting takes place, the errors saturate.
The solution difference for the method in~\cite{sundaram2018distributed} is plotted as $\epsilon_2^s$. In this algorithm, nodes communicate with their neighbors at all time steps, but only use those neighbors' data which are not among the extremes in the sorted set of neighborhood data points. Notice here that the malicious nodes might persist in the filtered neighborhood of regular nodes over time. Thus, this algorithm can only guarantee convergence in the convex hull of the minimizers of the regular nodes' private functions.

In Figure~\ref{fig:total_var}, we plot the errors produced by the algorithm proposed in~\cite[Algorithm 1]{ben-ameur2016robust}, where the authors replace the consensus constraint in the problem in~\eqref{eq:opt} with a TV norm regularizer to the objective, thereby penalizing nodes for not being in consensus with their neighbors. The problem is given by:
\begin{equation}\label{eq:TV-opt}
\begin{aligned}
& \min_{\bm{x}\in\mathbb{R}^d}~\frac{1}{|\mathcal{V}|} \sum_{i \in \mathcal{V}} f_i(\bm{x}_i) + \lambda\sum_{ij \in \mathcal{E}}(\bm{x}_i - \bm{x}_j),
\end{aligned}
\end{equation}
where $\lambda$ is the regularization parameter which controls the strength of magnitude of the penalty. If $\lambda=0$, consensus among the nodes is not enforced, thereby leading the nodes to their individual local minimizers, a node $i$ converges to the minimizer of $f_i$, $\forall i \in \mathcal{V}$. As $\lambda$ increases past zero, the consensus constraint is enforced, indicated by $\gamma_2$ reaching zero. While this method is successful in imposing consensus among the regular nodes, it can not, unlike RSGP, reduce the $\epsilon_p$ to zero.

\section{Conclusion}\label{sec:conclusion}

A robust decentralized optimization algorithm (RSGP) resilient to malicious Byzantine agents is proposed. This algorithm forgoes the need for the knowledge of out-degrees and works even for non-strongly convex loss functions. The algorithm dynamically isolates malicious nodes in the system thereby leading the system to convergence at the optimum consensus point. The isolation strategy is local to each node, is independent of the network topology and the attack strategy, and leverages the structure of the regular nodes in the system.


\bibliography{ref}
                                                     
\end{document}